\documentclass[pra,twocolumn,superscriptaddress]{revtex4-2}
\usepackage{amsmath,amssymb,amsfonts,bbm,graphicx,times,psfrag}
\usepackage[pdftex]{color}
\usepackage{amsmath,bm}
\usepackage[colorlinks, linkcolor=blue, citecolor=blue, urlcolor=blue, breaklinks=true]{hyperref}
\usepackage{microtype}

\usepackage{amsthm}
\usepackage{csquotes}
\usepackage{amsthm}\newcommand{\be} {\begin{equation}}
\newcommand{\ee} {\end{equation}}

\newcommand{\Tr}{{\rm Tr}}

\usepackage{color}
\renewcommand{\prl}{{Phys. Rev. Lett.} }
\renewcommand{\pra}{{Phys. Rev. A} }

\begin{document}
\title{Joint quantum estimation of loss and nonlinearity in driven-dissipative Kerr resonators}
\author{Muhammad Asjad}
\affiliation{Department of Applied Mathematics and Sciences, Khalifa University, 127788, Abu Dhabi, United Arab Emirates}
\author{Berihu Teklu}
\affiliation{Department of Applied Mathematics and Sciences, Khalifa University, 127788, Abu Dhabi, United Arab Emirates}
\affiliation{Center for
Cyber-Physical Systems (C2PS), Khalifa University, 127788, Abu Dhabi,
United Arab Emirates}
\author{Matteo G. A. Paris}
\affiliation{Quantum Mechanics Group \& Quantum Technology Lab, Dipartimento di Fisica 'Aldo Pontremoli dell'Universit\'a degli studi di Milano, I-20133 Milano, Italia}
\affiliation{Istituto Nazionale di Fisica Nucleare - Sezione di Milano, I-20133 Milano, Italia}
\date{\today}

\begin{abstract}
We address multiparameter quantum estimation for coherently driven nonlinear Kerr resonators in the presence of loss. In particular, we consider the realistic situation in which the parameters of interest are the loss rate and the nonlinear coupling, whereas the amplitude of the coherent driving is known and externally tunable.  Our results show that this driven-dissipative model  is asymptotically classical, i.e. the Uhlmann curvature vanishes, and the two parameters may be jointly estimated  without any additional noise of quantum origin. We also find that the ultimate bound to precision, as quantified by the quantum Fisher information (QFI), increases with the interaction time and the driving amplitude for both parameters. Finally, we investigate the performance of quadrature detection, and show that for both parameters the Fisher information oscillates in time, repeatedly approaching the corresponding QFI. 
\end{abstract}
\maketitle
\section{Introduction}
Quantum properties of radiation and matter may provide enhanced precision in metrological applications \cite{Giovannetti,Matteo}.  Indeed, quantum systems are notoriously 
fragile and very sensitive to  perturbations, even very weak perturbations, which makes them inherently  precise sensors.  Quantum probes have been exploited to improve sensitivity \cite{Degen,Pirandola,qp1,qp2,qp3,qp4,qp5}, and the resulting quantum technology represents a fundamental tool to improve metrological standards and increase 
precision of several characterization techniques ranging from stochastic noise \cite{qn1,qn2,qn3,qn4} to material science \cite{Marchiori,Thiel,Yip,Lesik,Hsieh}, 
and from biology \cite{Kucsko,Choi,Li,Marais} to gravitational wave detection \cite{McCuller,Abbott}.

For a single parameter of interest, the performance of a quantum sensor may be 
characterized in terms of the quantum Fisher information (QFI), which bounds the 
achievable precision via the quantum Cram\'{e}r-Rao bound (QCRB). On the other hand, in many realistic situations one is interested in the joint estimation of two ore more parameters \cite{Szc16,Carollo2018,Carollo2019,Liu19,Albarelli,Sho20,Candeloro}. Besides, the 
multi-parameter case has fundamental interest for studying compatibility 
problems involving the non-commutativity of quantum operations 
\cite{Hei16,Rag16,Bel21,Sid21}. In turn, the impossibility of jointly measuring non-compatible quantum observables makes it theoretically impossible to reach the multi-parameter version of the QCRB bound.

In this framework, driven-dissipative systems are of particular interest. In this kind of systems, one may explore the joint estimation of loss and nonlinearity by tuning the external driving in order to achieve the optimal working regime, possibly exploiting 
the existence of a stationary solution. In particular, here we address the paradigmatic example of driven-dissipative Kerr-resonator \cite{Drummond}, which represents a 
sensitive probe due to its strongly nonlinear response \cite{vm14,red16,Krimer19,Montenegro,zh21,sala21,li22,Bibak22,Kewming22,Xie22}. We investigate
situations where the parameters of interest are the strength of the Kerr nonlinearity $\chi$ and the rate of (one-photon) loss $\gamma$, which are estimated by probing the optical medium with suitable optical signals. We focus on 
the non equilibrium dynamics of the Kerr resonator, where the driving competes with incoherent dissipation and coherent amplification \cite{Sieberer,Rotter,Shovan}, and seek
for configurations where the joint estimation of loss and nonlinearity is possible without any additional noise of quantum origin, possibly enhancing the single-parameter 
sensitivity of the system.
 
The structure of the paper is as follows. In Section \ref{sec:2} we introduce the model, and in Section \ref{sec:qfi} we briefly review the basic tools of local 
multi-parameter quantum estimation theory. Specifically, we outline the 
ultimate bounds to precision in the joint estimation of the driven-dissipative Kerr nonlinear resonator. In Section \ref{sec:res}, we illustrate and discuss our results: 
in Section \ref{sec:resi} we show results about the time evolution of the quantum Fisher information matrix, and in Section \ref{sec:resh} we discuss the performance of homodyne detection in the estimation of the two parameters.
Finally, Section \ref{sec:con} closes the paper with some concluding remarks.

\section{Preliminaries}
\label{sec:bigsec2}
\subsection{The driven-dissipative Kerr resonator}
\label{sec:2}
The system we study is a general model of a single driven-dissipative Kerr nonlinear resonator of  frequency $\omega_c$ whose coherent dynamics is described by the Hamiltonian
\begin{equation}
H=-\Delta \hat{a}^{\dagger}\hat{a}+\frac{\chi}{2} \hat{a}^{\dagger}\hat{a}^{\dagger}\hat{a} \hat{a}-i F(\hat{a}-\hat{a}^{\dagger})\,, \label{h1}
\end{equation}
in the rotating-wave approximation, where $\Delta=\omega_{p}-\omega_{c}$ is the pump-cavity detuning, $F$ is the coherent drive strength, and $\chi$ is the Kerr anharmonicity. The operator $\hat{a}^{\dagger}(\hat{a})$ is creation (annihilation) operator of the resonator. We assume that the resonator is coupled to a zero-temperature bath. 
Therefore, the full dissipative dynamics of such a system are described by the Lindblad master equation 
\begin{equation}
  \frac{d}{dt} \hat{\rho}=-i[\hat{H},\hat {\rho}(t)]+\frac{\gamma}{2} \mathcal{D} [\hat{a}] \hat{\rho} \equiv 
        \mathcal{L}_0  \hat{\rho}, \label{eqm}
\end{equation}
where $\mathcal{D}[\hat{X}] \hat \rho \equiv \hat{X} \hat{\rho} \hat{X}^\dagger - (1/2) \left \{ \hat{X}^\dagger \hat{X}, \hat{\rho} \right \}$  is the usual Lindblad dissipative superoperator, and $\gamma$ is the one photon decay rate. 
Systems described by the Hamiltonian in Eq. (\ref{h1}) may be realized using several experimental platforms such as semiconductor microcavities \cite{,Malpuech,Rodriguez,Fink,Geng}, quantum circuits \cite{Brookes,Leghtas,Blais}, and 
optomechanical setups \cite{Benito, Sanchez}.

\subsection{Multiparameter quantum estimation} \label{sec:qfi}
The quantum Fisher information is the main tools of quantum sensing and metrology, 
as its quantifies the ultimate precision in estimating a parameter encoded onto a 
quantum state. In turn, it also serves to assess whether and how a certain quantum 
system may be exploited as a sensing device. Let us consider a scenario in which we are interested in estimating the value of a vector 
of parameters, ${\bm\lambda}=(\lambda_1,\lambda_2,...)$ encoded onto a quantum state \begin{equation}
\rho({\bm \lambda})=\sum_{k}p_{k}({\bm \lambda}) |\rho_{k}({\bm \lambda})\rangle\langle p_{k}({\bm \lambda})| \,.
\end{equation}
The quantum Fisher information matrix (QFIM) is defined as 
\begin{equation}\label{eq:Fisher}
\mathcal{F}_{jk}=\frac{1}{2}
\, {\rm Tr}
\big[\rho_{\lambda}\,\{{\cal L}_j,{\cal L}_k\}\big]\,, \end{equation}
where $\{A,B\}=A B+ B A$ denotes the anticommutator, and the operators ${\cal L}_j\equiv {\cal L}_{\lambda_j}$ are the so-called symmetric logarithmic derivatives with respect to the parameter $\lambda_j$, which are defined implicitly by the relations
\begin{equation}
2\partial_{\lambda_j} \rho =\rho {\cal L}_{j}+{\cal L}_j \rho\,.
\end{equation}  
The QFIM provides a matrix lower bound on the average mean-square error (MSE) matrix 
of the estimates, usually referred to as the multiparameter quantum Cram\'er-Rao bound (QCRB)
\begin{equation}
\text{\bf V}(\bm{\hat \lambda};\bm\lambda)\ge \frac{1}{M}\mathcal{F(\bm\lambda)}^{-1},
\end{equation}
where $\text{\bf V}(\bm{\hat \lambda};\bm\lambda)$ is the covariance matrix of  
${\bm\lambda}$, $\text {M}$ the number of independently repeated measurements.

If we now introduce the $d\times d$ real, weight matrix $W$ (e.g. the identity matrix),  
we may obtain a useful scalar bound as
\begin{equation}
\label{eq:SLDQFIboundscalar}
\Tr \left[\bm{W}\bm{V}(\bm{\hat{\lambda}},\bm\lambda)\right] \geq \Tr \left[ \bm{W}{\bm{\mathcal{F}}}^{-1} (\bm{\lambda}) \right] =
C_S(\bm{W},\bm\lambda)\,,
\end{equation}
which is usually referred to as the  SLD-QFI scalar bound, and represents a 
benchmark  for multiparameter estimation. The SLD-QCRB is generally not attainable, 
due to the incompatibility of generators of different parameters, which is reflected by the non commutativity of the corresponding SLDs. Indeed, the optimal 
measurement operators corresponding for the different parameters may 
not commute with each other, making this scalar bound unreachable. 
The SLD-QRCB is achievable if the Uhlmann curvature matrix with elements 
\cite{Carollo2018} 
\begin{equation}
\big[ \bm{\mathcal{U}}(\bm\lambda) \big]_{nm} = -\frac{i}{2}\Tr\left[\rho_{\bm\lambda} \left[{\cal L}_n,{\cal L}_m \right]\right]\,,
\end{equation}
vanishes. We recall that multiparameter quantum metrology corresponds to simultaneous estimation of multiple parameters using a single quantum 
system to probe a quantum dynamics with unknown parameters. In other words, 
if one wishes to estimate separate parameters as precisely as one would 
estimate them individually when assuming perfect knowledge of the other 
parameters, then a compatibility conditions need to be 
satisfied, which is precisely the vanishing of the Uhlmann curvature matrix $\bm{\mathcal{U}}(\bm\lambda) = \bm{0}$. This ensures the 
existence of compatible measurements and the possibility of saturating the SLD-QCRB.
In addition, there must exist a single probe state $\rho_{\lambda}$ leading
to the optimal QFI for each of the parameters under consideration.  

In our case, we have two parameters, $\chi$ and $\gamma$ and the corresponding SLDs are defined implicitly (at any time) by 
\begin{align}
2\partial_{\chi} \rho(t)& =\rho(t)\mathcal{L}_{\chi}+\mathcal{L}_{\chi}\rho(t) \\
2\partial_{\gamma} \rho(t)& =\rho(t)\mathcal{L}_{\gamma}+\mathcal{L}_{\gamma}\rho(t)
\end{align}  
In order to evaluate the QFIM and the Uhlmann matrix at any time we write 
the derivative of the master equation (\ref{eqm}) with respect to the parameter 
$\chi$ and $\gamma$ and obtain:
\begin{align}
\frac{d}{dt} \partial_\chi\hat{\rho}=&-i[\hat{H},\partial_\chi\hat {\rho}(t)]-i[\partial_\chi\hat{ H},\hat {\rho}(t)]\\ \notag
&+\frac{\gamma}{2} \mathcal{D} [\hat{a}] \partial_\chi\hat{\rho} \equiv \partial_\chi (\mathcal{L}_0  \hat{\rho})\\ \notag
\frac{d}{dt} \partial_\gamma\hat{\rho}=&-i[\hat{H},\partial_\gamma\hat {\rho}(t)] +\frac{1}{2}\mathcal{D} [\hat{a}] \hat{\rho}\\ \notag&+  \frac{\gamma}{2}  \mathcal{D} [\hat{a}] \partial_\gamma \hat{\rho} \equiv \partial_\gamma(\mathcal{L}_0  \hat{\rho}) \notag
\end{align}
where we have used the notation $\partial_k\equiv\partial/\partial k$ for $k=(\chi,\gamma$). 
The above set of equations, together with the master equation (\ref{eqm}) 
can be written in a compact matrix form as
$\dot{\mathcal{R}}(t)=\mathcal{A} \mathcal{R}(t)$,  where $\mathcal{R}(t)=[\rho(t),\partial_\chi\rho(t), \partial_\gamma \rho(t)]^T$ and $\mathcal{A}$ takes the form
 \begin{equation}\label{system}
\mathcal{A}=\left(\begin{array}{ccc}
\mathcal{L}_0 &                           0&                    0\\
\partial_\chi\mathcal{L}_0&      \mathcal{L}_0&             0  \\
\partial_\gamma\mathcal{L}_0&      0&             \mathcal{L}_0   \\
\end{array}\right)
\end{equation}
We end this Section by reminding that if the Uhlmann curvature vanishes, then it makes sense to address single-parameter estimation. In this case the relevant CR bounds 
are given by
\begin{align}
\hbox{Var}\lambda \geq \frac{1}{M {\bm{\mathcal{F}}}_{\lambda\lambda}}
\end{align}
where $\lambda=\chi,\gamma$, $\hbox{Var}$ denotes the variance of any unbiased estimator, and the quantities ${\bm{\mathcal{F}}}_{\lambda\lambda}$ are the diagonal elements of the QFIM. These elements also provide bounds to the classical Fisher information of any observable aimed at estimating the parameter $\lambda$ as follows. If we perform a measurement described by the probability operator-valued measure (POVM) $\left\{\Pi_x\right\}$, with $\sum \Pi_x = {\Bbb I}$, and obtain the distribution of outcomes $p(x|\lambda) = \hbox{Tr}\left[\rho_\lambda\,\Pi_x\right]$, then we have ${\mathcal F}_X(\lambda) \leq {\bm{\mathcal{F}}}_{\lambda\lambda}$ where ${\mathcal F}_X(\lambda) = \sum_x \left[\partial_\lambda\,p(x|\lambda) \right]^2/p(x|\lambda)$ is the Fisher information of $p(x|\lambda)$ i.e. the maximum information about $\lambda$ that may extracted by measuring $\left\{\Pi_x\right\}$.
\begin{figure*}[t!] 
\begin{center}
\includegraphics[width=0.99 \textwidth]{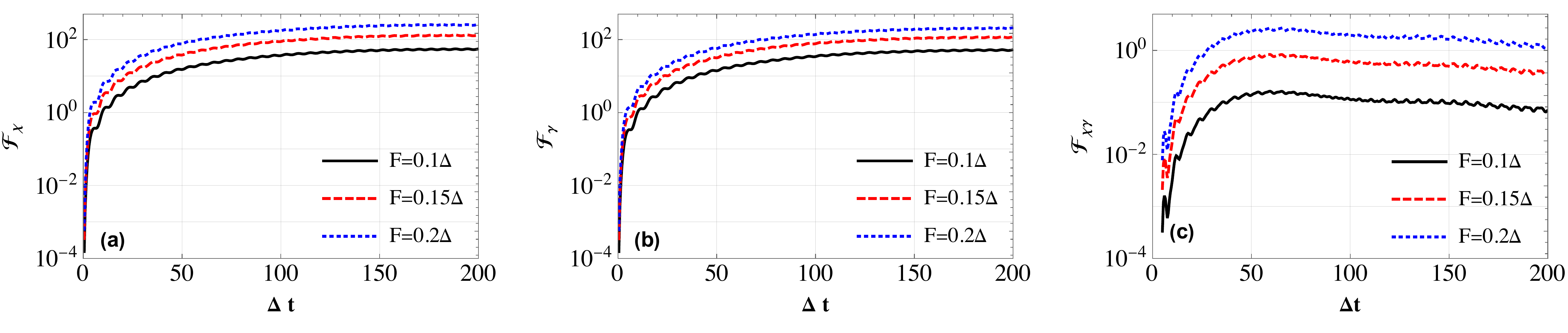} \\
\includegraphics[width=0.99 \textwidth]{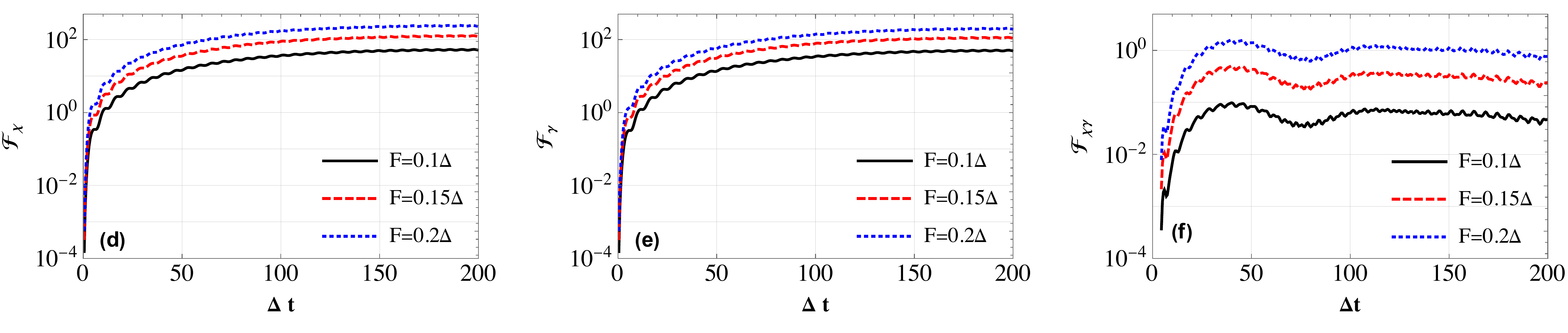} \\
\includegraphics[width=0.99 \textwidth]{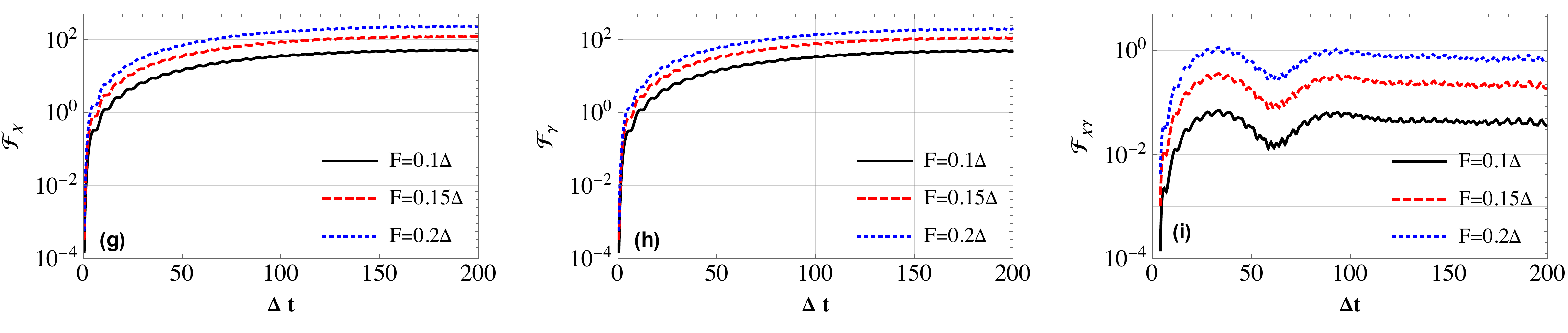}
\caption{The quantum Fisher information matrix elements as a function of (renormalized) evolution time $\Delta t$. The first column show results for ${\cal F}_\chi$, the second one for ${\cal F}_\gamma$, and the last one shows  the off-diagonal element of the QFIM 
${\cal F}_{\chi\gamma}$. Parameters are set to $\chi=0.1\Delta$ (first row),$\chi=0.5\Delta$ (second row) and $\chi=\Delta$ (third row) and the different curvs refer to different values of the driving strength $F=0.01\Delta$ (black curves), $F=0.1\Delta$ (blue curves) and $F=\Delta$ (red curves). We also set $\gamma=0.01\Delta$. \label{f:1}}    
\end{center}
\end{figure*}
\section{Results}
\label{sec:res}
\subsection{The elements of the QFIM and the Uhlmann curvature}
\label{sec:resi}
Upon solving the system of coupled differential equations in (\ref{system}), we may obtain the SLDs for the two parameters $\chi$ and $\gamma$ and, in turn, the QFIM. In general, it is challenging to obtain symmetric logarithmic derivative $\mathcal{L}_{\chi,\gamma}$ analytically. In our case,  we solve the above set of equations numerically to obtain the SLDs and then QFIM. In doing this, we are free to tune the value of the driving $F$ and the detuning $\Delta $, since this degrees of freedom are typically available in realistic situations. In particular, we evaluate the elements of the QFIM as a function of the renormalized interaction time $\Delta t$ (since varying $\Delta$ simply corresponds to a change of the overall timescale) and for different values of the driving. 

In Figs. 1 (a-c), we show the typical behaviour of the QFI elements as a 
function of $\Delta t$ for different values of the coherent drive strength $F$. 
The specific values used for these plots are $\chi=0.1\Delta$ and 
$\gamma=0.01\Delta$. We use the notation ${\cal F}_\lambda \equiv {\cal F}_{\lambda\lambda}$ for the diagonal elements of the QFIM, $\lambda=\chi,\gamma$ and ${\cal F}_{\chi \gamma}$ for the off-diagonal one. For small times, all the elements of the QFIM increase quite rapidly and then tend to saturate, at least in the range of interaction times we have explored numerically. Notice that our system is known to possess a stationary state \cite{ciuti}, and this corroborates our findings, indicating the saturation of the values of the elements of the QFIM.

The off-diagonal term of the QFIM is nearly zero in all the conditions we have investigate numerically and, in turn, the 
Uhlmann curvature vanishes. This means that the loss and nonlinearity parameters may be jointly estimated, and no additional noise of quantum origin occurs in inferring 
them from measured data. 

The diagonal terms of the QFIM govern the ultimate precision achievable in estimating $\chi$ and $\gamma$. Both elements increases with time and the amplitude of the coherent driving. A similar behaviour may be observed for other values of the parameters. Overall, we conclude that joint estimation of loss and nonlinearity is possible, and to maximize precision one has to employ a moderately large driving (in unit of $\Delta$) and a large interaction time ($\Delta\, t \sim 100$).

\begin{figure*}[t!]
\centerline{\includegraphics[width=0.95\textwidth]{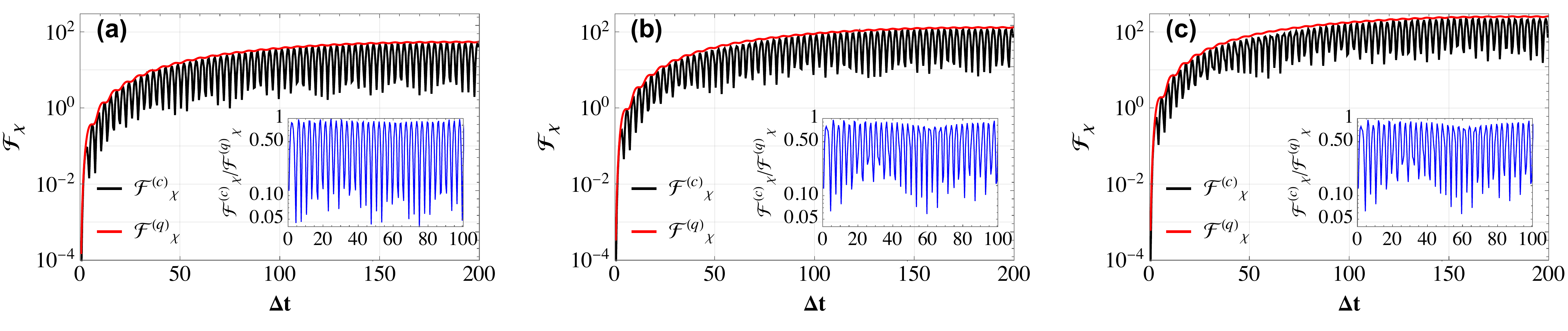}}
\caption{Homodyne Fisher information $\mathcal{F}_X(\chi)$ for the nonlinearity 
as function of the rescaled time $\Delta t$ for $F=0.1 \Delta$(a), $F=0.15 \Delta$ (b) and  $F=0.2 \Delta$ (c) for fixed value of $\chi=0.05\Delta$. The corresponding diagonal element 
of the QFIM $\mathcal{F}_\chi$ is shown for comparison.
The insets show the ratio $\mathcal{F}_X(\chi)/\mathcal{F}_\chi$. 
The other parameters are set as in Fig. 1.\label{f:2}}     
\end{figure*}   

\begin{figure*}[t!]
\includegraphics[width=0.95\textwidth]{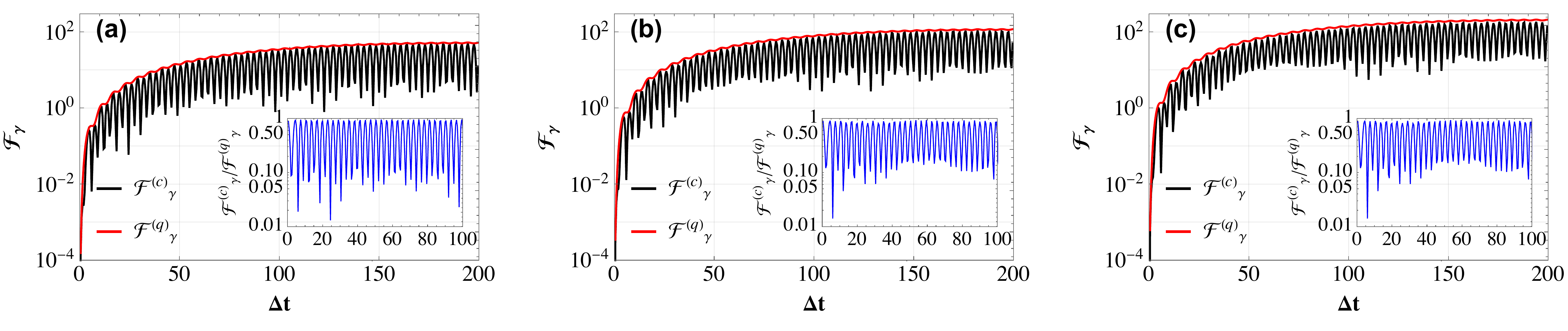} 
\caption{Homodyne Fisher information $\mathcal{F}_X(\gamma)$ for the loss 
as function of the rescaled time $\Delta t$ for $F=0.1 \Delta$(a), $F=0.15 \Delta$ (b) and  $F=0.2 \Delta$ (c) for fixed value of $\gamma=0.01\Delta$. The corresponding diagonal element of the QFIM $\mathcal{F}_\gamma$ is shown for comparison.
The insets show the ratio $\mathcal{F}_X(\gamma)/\mathcal{F}_\gamma$. 
The other parameters are set as in Fig. 1.\label{f:3}}            
\end{figure*}

\subsection{Estimation in practice: performance of homodyne detection}
\label{sec:resh}

In this section, we analyze the performance of homodyne detection, i.e. we evaluate the classical Fisher information of this detection scheme for one of the parameter and compare its value with the corresponding (diagonal) element of the QFIM. Homodyne detection measure a quadrature of the field, i.e. the Hermitian operator
\begin{equation}
x_\theta=\frac{1}{2}(a e^{-i \theta}+a^{\dagger}e^{i\theta})\,.
\end{equation}  
In our case, we set $\theta$ to zero. The POVM of the detector is given by $\Pi_x = |x\rangle\langle x|$ where, in the Fock representation,  $|x\rangle$ is given by
\begin{equation}
|x \rangle=  \dfrac{1}{\pi^{1/4}}\sum_n   \dfrac{ \exp(-x^2/2)}{2^{n/2} (n!)^{1/2}}   \,   H_n(x)\, |n\rangle, 
\end{equation}
where $H_{n}(x)$ is the Hermite polynomial of order $n$. Since the quadrature has a continuous spectrum, the classical Fisher information is 
given by 
\begin{equation}\label{eq:FC}
\mathcal{F}_X(\lambda)=\int dx \frac{1}{p(x|\lambda)}\bigg[ \frac{\partial p(x|\lambda)}{\partial\lambda}\bigg]^2,
\end{equation}  
where $p(x|\lambda) = \hbox{Tr}\left[\rho_\lambda\, \Pi_x\right]$. 

By using the numerical solution of the master equation (\ref{eqm}), we evaluate the distribution $p(x|\chi,\lambda)$ at any time and, in turn, the two classical Fisher informations $\mathcal{F}_X(\chi)$ and $\mathcal{F}_X(\gamma)$. Results are illustrated in Figs. \ref{f:2} and \ref{f:3}

In Figs. 2 (a-c), we show $\mathcal{F}_X(\chi)$ as a function of the rescaled time $\Delta t$ together with the QFIM element   $\mathcal{F}_\chi$ for $\chi=0.05\Delta$ and different values of the driving amplitude $F$. As it is apparent from the plots, the homodyne Fisher 
information oscillates in time, with the maxima being very close to $\mathcal{F}_\chi$. In other words, homodyne detection provides a nearly optimal estimation scheme for the nonlinearity, though an accurate selection of the measurement time is required. In the insets of Figs. 2 (a-c), we show the ratio $\mathcal{F}_X(\chi)/\mathcal{F}_\chi$, which confirms the observations made above. Notice that both, $\mathcal{F}_X(\chi)$ and 
$\mathcal{F}_\chi$ increase with the amplitude of the coherent driving $F$.

Figs. 3 (a-c) contains results similar to those of in Figs. 2 (a-c), but for the loss parameter $\gamma$, i.e. $\mathcal{F}_X(\gamma)$ as a function of the rescaled time $\Delta t$ together with the QFIM element   $\mathcal{F}_\gamma$ for $\gamma=0.01\Delta$ and different values of the driving amplitude $F$. Also for the loss parameter, 
the homodyne Fisher information oscillates in time, with the maxima being very close to $\mathcal{F}_\gamma$, i.e. homodyne detection is a nearly optimal estimation scheme \cite{pax00} provided that an accurate selection of the measurement time is avalaible. In the insets, we show the ratio $\mathcal{F}_X(\gamma)/\mathcal{F}_\gamma$.

Notice that by changing the phase of the measured quadrature, we may change the position of the maxima of the homodyne Fisher information, i.e. a feedback mechanism may be used to achieve optimality.

\section{Conclusions} \label{sec:con}
In this paper, we have addressed the joint estimation of loss and nonlinearity in driven-dissipative Kerr resonators, i.e. coherently driven Kerr oscillators in the presence of loss. In particular, we have considered the realistic situation in which the loss rate and the nonlinear coupling are the parameter of interest, whereas the amplitude of the coherent driving, and the detuning, are known and externally tunable. 

Our results show that this driven-dissipative model is asymptotically classical, 
i.e. the Uhlmann curvature vanishes, and the two parameters may be jointly estimated without any additional noise of quantum origin. We have also found that the ultimate 
precision, as quantified by the quantum Fisher information (QFI), may be improved, for both parameters, by increasing the interaction time and the driving amplitude. 

Finally, we have investigated the performance of quadrature measurements, i.e. homodyne detection, and have shown that for both parameters homodyne Fisher information oscillates in time, repeatedly approaching the corresponding QFI. In other words, homodyne detection provides a nearly optimal estimation scheme for both loss and nonlinearity, though an accurate selection of the measurement time is required in both cases. 

Our results show that driven-dissipative Kerr resonators are convenient probes for jointly estimating nonlinearity and loss without additional quantum noise, and pave the way for 
the experimental implementaton in quantum optical systems.

\section*{Acknowledgments}
\noindent This work has been supported by Khalifa University of Science and 
Technology under Award No. FSU-2021-018 (Estimation of Nonlinearities in Optical Media). This work was done under the auspices of INdAM-GNFM. MGAP thanks R. F. Antoni for motivating comments.

\end{document}